\documentclass[preprint]{aa}
\usepackage{graphicx}
\usepackage[varg]{txfonts}
\usepackage{cases}
\usepackage{natbib}
\usepackage{underscore}
\usepackage{hyperref} 

\defcitealias{PerriLeitner2022}{Perri \& Leitner et al.} 
\defcitealias{Mikic1999}{Miki\'c et al., 1999} 
\defcitealias{Kuzma2023}{Ku\'zma et al., 2023} 

\begin{document}
      \title{How coronal mass ejections are influenced by the morphology and toroidal flux of their source magnetic flux ropes?}
      
      \author{J. H. Guo\href{https://orcid.org/0000-0002-4205-5566}{\includegraphics[scale=0.05]{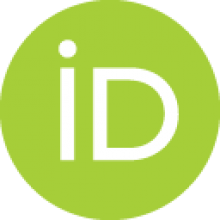}}\inst{1,2}, L. Linan\inst{1}, S. Poedts\inst{1,3}, Y. Guo\inst{2}, B. Schmieder\inst{1,4}, A. Lani\inst{1}, Y. W. Ni\inst{2}, M. Brchnelova\inst{1}, B. Perri\inst{5,1}, T. Baratashvili\inst{1}, S. T. Li\inst{6,7}, P. F. Chen\inst{2}}
      
      \institute{Centre for mathematical Plasma Astrophysics, Department of Mathematics, KU Leuven, Celestijnenlaan 200B, B-3001 Leuven, Belgium 
        \email{Stefaan.Poedts@kuleuven.be}
        \and
        School of Astronomy and Space Science and Key Laboratory of Modern Astronomy and Astrophysics, Nanjing University, Nanjing 210023, China 
      	\email{chenpf@nju.edu.cn}
        \and
        Institute of Physics, University of Maria Curie-Skłodowska, ul.\ Marii Curie-Sk{\l}odowskiej 1, 20-031 Lublin, Poland
        \and
        LESIA, Observatoire de Paris, CNRS, UPMC, Universit\'{e} Paris Diderot, 5 place Jules Janssen, 92190 Meudon, France
        \and
        AIM/DAp - CEA Paris-Saclay, Université Paris-Saclay, Université Paris-Cité, Gif-sur-Yvette, France
        \and
        Key Laboratory of Dark Matter and Space Astronomy, Purple Mountain Observatory, Chinese Academy of Sciences, Nanjing 210023, People’s Republic of China
        \and
        INAF-Turin Astrophysical Observatory, Via Osservatorio 20, 10025, Pino Torinese (TO), Italy
 	    }
    \titlerunning{Deducing the CMEs from their progenitors in the solar source region}
 \authorrunning{Guo et al.}
\date{}

\abstract
{Coronal mass ejections (CMEs) stand as intense eruptions of magnetized plasma from the Sun, playing a pivotal role in driving significant changes of the heliospheric environment. Deducing the properties of CMEs from their progenitors in solar source regions is crucial for space weather forecasting.}
{The primary objective of this paper is to establish a connection between CMEs and their progenitors in solar source regions, enabling us to infer the magnetic structures of CMEs before their full development.} 
{We create a dataset comprising a magnetic flux rope series with varying projection shapes (S-, Z- and toroid-shaped), sizes and toroidal fluxes, using the Regularized Biot-Savart Laws (RBSL). These flux ropes are inserted into solar quiet regions, aimed at imitating the eruptions of quiescent filaments. Thereafter, we simulate the propagation of these flux ropes from the solar surface to a distance of 25$R_{\odot}$ with our global coronal magnetohydrodynamic (MHD) model which is named COCONUT.}
{Our parametric survey reveals significant impacts of source flux ropes on the consequent CMEs. Regarding the flux-rope morphology, we find that the projection shape (e.g., sigmoid or torus) can influence the magnetic structures of CMEs at 20$R_{\odot}$, albeit with minimal impacts on the propagation speed. However, these impacts diminish as source flux ropes become fat. In terms of toroidal flux, our simulation results demonstrate a pronounced correlation with the propagation speed of CMEs, as well as the successfulness in erupting.}
{This work builds the bridge between the CMEs in the outer corona and their progenitors in solar source regions. Our parametric survey suggests that the projection shape, cross-section radius and toroidal flux of source flux ropes are crucial parameters in predicting magnetic structures and propagation speed of CMEs, providing valuable insights for space weather prediction. On the one hand, the conclusion drawn here could be instructive in identifying the high-risk eruptions with the potential to induce stronger geomagnetic effects ($B_{z}$ and propagation speed). On the other hand, our findings hold practical significance for refining the parameter settings of launched CMEs at 21.5$R_{\odot}$ in heliospheric simulations, such as with EUHFORIA, based on observations for their progenitors in solar source regions.}

\maketitle

\section{Introduction}\label{introduction} 

Coronal mass ejections (CMEs) represent the most intense and powerful explosions in the solar system, standing as primary drivers of changes in the solar-terrestrial space environment \citep{Zhang2007, Chen2011, Richardson2012, Schmieder2015, Kilpua2017b}. When they reach the Earth, their southward magnetic fields can induce magnetic reconnection with the intrinsic geomagnetic field, thereby injecting energy and energetic particles from the solar wind into the magnetosphere. This has the potential to jeopardize satellites and even pose risks to human health \citep{Gosling1987, Schrijver2015}. Hence, predicting the properties of CMEs, particularly the magnetic structures (like the $B_{z}$ profile) and propagation speed before their full development, remains a priority for space weather prediction.

Extensive observations have revealed that the core magnetic structure of CMEs is a magnetic flux rope, i.e.\ bundles of twisted field lines winding around a common axis \citep{Schmieder2006, Cheng2017, Patsourakos2020, Liu2020}. White-light coronagraph observations suggested that at least 40\% of CMEs may incorporate a flux rope \citep{Vourlidas2013}, manifested as a dark cavity or bright core in observations \citep{Chen2011, Song2022}. Furthermore, in-situ detections of their consequences in interplanetary space, such as magnetic clouds, indicate that they often exhibit smooth and large-angle rotations in vector magnetic fields, suggesting the presence of helical field lines \citep{Burlaga1981, Klein1982}.  

Although the consensus regarding flux ropes as the magnetic cores of CMEs has been well accepted, the magnetic configuration of their progenitors remains elusive: whether a magnetic flux rope exists before the eruption or forms during it? Addressing this issue requires investigations on the magnetic structure of CME progenitors, such as hot channels \citep{Zhang2012, Cheng2017}, filaments \citep{Schmieder2013, Chen2020} and cavities \citep{Gibson2015}. For example, by diagnosing the magnetic configuration of eruptive filaments, \citet{Ouyang2015} demonstrated that flux ropes are not a necessary condition for CME progenitors. Similarly, \citet{Song2014} and \citet{Wangws2017} found that flux ropes can form via magnetic reconnection during CME propagation but not before. Nonetheless, various observations and numerical modeling studies demonstrated that flux ropes exist before the eruption in many events. For instance, certain observed proxies before the flare onset, such as conjugate coronal dimmings \citep{Webb2000, Qiu2007, Wang2019, Xing2020, Wang2023} and sunspot scar \citep{Xing2023}, have been identified as the footpoints of pre-eruptive flux ropes. Additionally, numerous non-linear-force-free-field (NLFFF) extrapolations and data-driven MHD simulations of coronal magnetic fields have also indicated the existence of preexisting flux ropes \citep{Cheung2012, Guo2010, Jiang2018, James2018, Duan2019, Guo2019a, Kilpua2019, Guojh2021, Inoue2021, Guojh2023a, Guo2024}. Beyond case studies, the percentage of different pre-eruptive magnetic structures of CME progenitors has been quantified. \citet{Ouyang2017} investigated the magnetic configuration of 576 eruptive filaments and found that approximately 89\% of them are supported by flux ropes. \citet{Wang2023} found that 9 events display pre-eruptive coronal dimmings within a dataset of 28 CME events associated with dimmings. This implies that, even though not necessary, many CMEs result from the eruption of magnetic flux ropes.

In addition to that, numerous studies have demonstrated the feasibility of deducing the properties of CMEs in the outer corona and interplanetary space from their solar progenitors. For example, many works revealed the consistency of the magnetic structure, such as the magnetic-field orientation and chirality, between interplanetary magnetic clouds and their progenitor filaments \citep{Bothmer1994, Marubashi1997, Yurchyshyn2001, Wang2006, Rodriguez2008}. In these cases, CME flux ropes largely maintain their initial orientation throughout the propagation process. However, there also exist CME flux ropes that undergo significant rotation in the eruption process, with the rotation direction strongly dependent on their projection shapes \citep{Green2007, Zhou2020, Zhou2022}. Besides, a substantial amount of magnetic helicity carried by interplanetary magnetic clouds is inherited from their precursors in solar source regions. \citet{Xing2020} discovered that toroidal fluxes which are contributed by preexisting flux ropes to magnetic clouds lie in a range of 40\% to 88\%. Through a comparison between the NLFFF extrapolation and reconstruction with in-situ measurement of a magnetic cloud, \citet{Thalmann2023} found that toroidal flux of the twisted magnetic structure in the solar source region can contribute 50\% of that of its consequent magnetic cloud. These findings unveil the intrinsic linkages between preexisting flux ropes and their consequent CMEs, indicating the possibility of predicting magnetic structures and dynamics of CMEs from remote-sensing observations in solar source regions, even prior to the onset of eruptions.

The main objective of this paper is to build a bridge between source magnetic flux ropes and their consequent CMEs. To this end, we construct a series of flux ropes with different morphology, size, and toroidal flux (intense of the electric current). Subsequently, we simulate the self-consistent propagation processes of these flux ropes from the solar surface to 25$R_{\odot}$, using a three-dimensional (3D) global coronal model, called COolfluid COroNal UnsTructured (COCONUT; \citetalias{PerriLeitner2022}~\citeyear{PerriLeitner2022}), as realized in \citet{Linan2023} and \citet{Guojh2024b}. This paper is organized as follows. We describe the methodology in Section~\ref{sec:met}, exhibit the results in Section~\ref{sec:res}, which are followed by a discussion in Section~\ref{sec:dis}. Finally, we summarize the findings in Section~\ref{sec:sum}.

\section{Methodology}\label{sec:met}

\subsection{Reconstruction of preexisting magnetic flux ropes}

To assess the impact of preexisting flux ropes in solar source regions on their consequent CMEs, it is crucial to construct flux ropes with high flexibility. For example, when investigating the influence of flux-rope morphology, it is advisable to allow the axes of flux ropes to be arbitrary rather than fixed. Moreover, the presence of conjugate coronal dimmings in observations suggests that eruptive flux ropes should be anchored to the solar surface rather than fully detached from it \citep{Wangws2017, Aulanier2019, Xing2020b, Wang2023}. Given these observational constraints, \citet{Titov2018} proposed a novel method to construct magnetic fields including a thin flux rope, called Regularized Biot-Savart Laws (RBSL). This method can construct a force-free flux rope with an axis following an arbitrary path, facilitating the investigation of the morphology of flux ropes. Recently, \citet{Guojh2024b} has implemented this technique in COCONUT and demonstrated its efficiency in modeling the initiation and propagation of a CME resulting from a sigmoid. Here, we will explore the impacts of the parameters associated with the RBSL flux rope, particularly its morphology and toroidal flux, on the consequent CMEs using COCONUT.

\begin{figure}
  \includegraphics[width=8cm,clip]{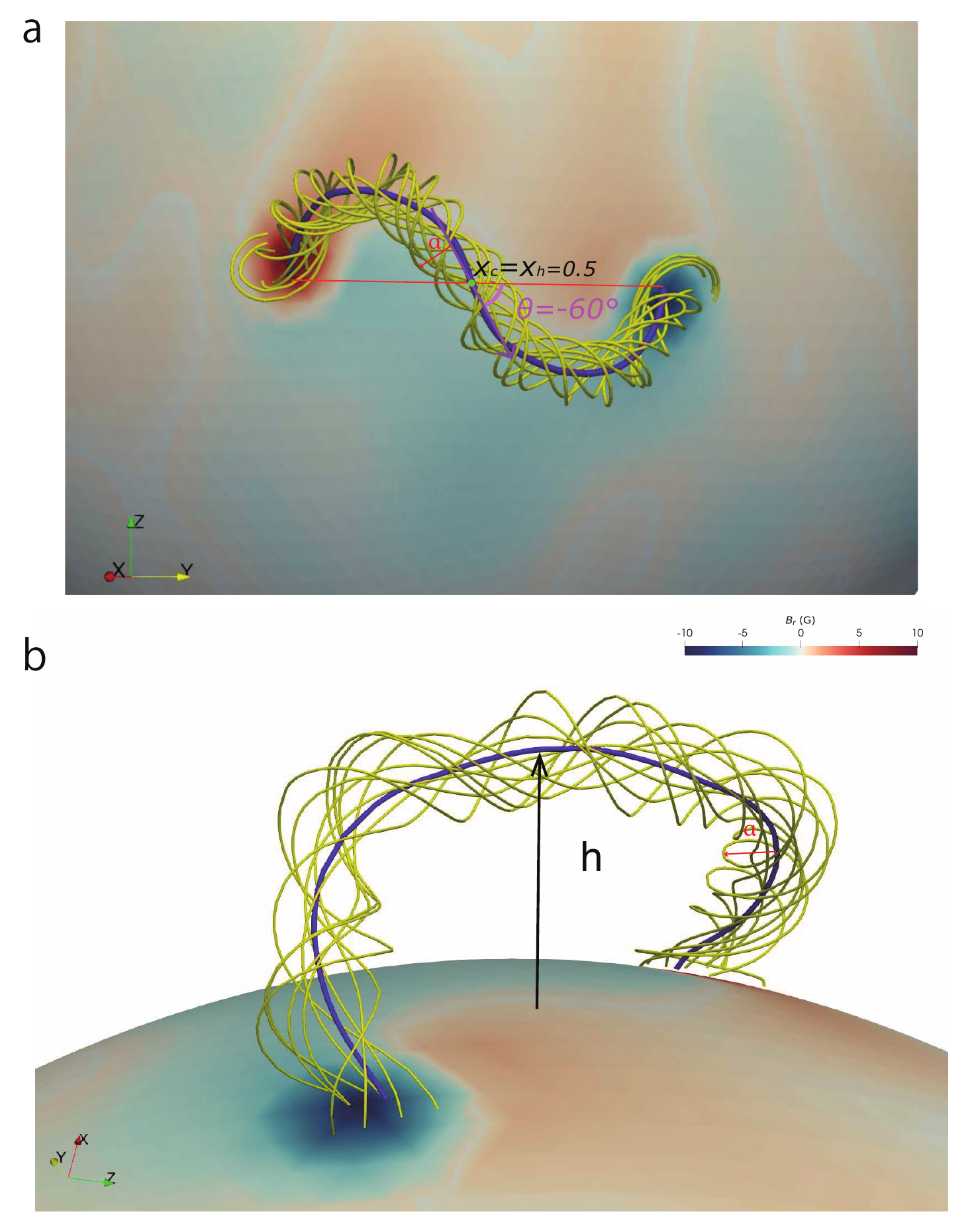}
  \centering
  \caption{Sketch of a typical Z-shaped magnetic flux rope with marked parameters of the RBSL flux rope, in which the purple line represents the flux-rope axis, and yellow lines represent twisted field lines. \label{sketch}}
\end{figure}

The reconstruction of RBSL flux ropes involves four primary parameters: the axis path ($C$), cross-section radius ($a$), toroidal flux ($F$), and electric current ($I$). Among these, the axis path and cross-section radius jointly determine the geometry of a flux rope, in which the former controls its projection shape (S-, Z- or toroid-shaped), while the latter governs its overall size (slender or fat). The toroidal flux $F$ influences both the magnetic-field strength and the electric current flowing through the flux rope ($F=\pm(3 \mu_{0} Ia)/5\sqrt{2}$, deduced from the equilibrium condition in the RBSL regime). To precisely control the flux rope morphology, similar to \citet{Guojh2024b}, the path of its axis is described by a theoretical curve which is defined as follows: 

 \begin{eqnarray}
 && f(s)=  \begin{cases} \frac{s(2x_{c}-s)}{x_{c}^{2}}\theta, \quad 0 \le s \le x_{\rm c} \\[8pt] \frac{(s-2x_{_{\rm c}}+1)(1-s)}{(1-x_{c})^{2}}\theta, \qquad x_{_{c}} < s \le 1 \end{cases} \label{eq12}
  \end{eqnarray}
\begin{eqnarray}
   &&x=(s-x_{c})\cos f + x_{c}, \label{eq13}\\
   &&y=(s-x_{c})\sin f, \label{eq14}
 \end{eqnarray}
  \begin{eqnarray}
 && z(x)=  \begin{cases} \frac{x(2x_{\rm h}-x)}{x_{\rm h}^{2}}h, \quad 0 \le x \le x_{\rm h} \\[8pt] \frac{(x-2x_{_{\rm h}}+1)(1-x)}{(1-x_{\rm h})^{2}}h, \qquad x_{_{\rm h}} < x \le 1 \end{cases} \label{eq15}
 \end{eqnarray}
 where $\theta$ determines the projection angle with respect to the line connecting two footpoints, $x_{c}$ controls the position where the projected curve intersects the line connecting two footpoints (referred to as the crossing point), $x_{h}$ represents the apex position, and $h$ indicates the apex height. Note that the sign of $\theta$ determines the projection shape (sign of non-local writhe) of the flux rope. Flux ropes which are constructed with values of $\theta>0$, $\theta<0$ and $\theta=0$ correspond to S-shaped, Z-shaped and toroid-shaped morphologies, respectively. Figure~\ref{sketch} illustrates the configuration of the RBSL flux rope ($\theta=-60^{\circ}$) with the key parameters marked, correspinding to a Z-shaped sigmoid.

In our parametric survey, flux ropes are positioned at the solar equator with fixed values of $x_{c}=x_{h}=0.5$ and an apex height of $h=120$~Mm. The parameters of our modeled flux ropes are in accord with large-scale quiescent filaments in observations \citep{Tandberg1995}. Figure~\ref{fig1} displays the flux ropes constructed with $\theta$ values of $60^{\circ}$ (Figures~\ref{fig1}a and \ref{fig1}b), $0^{\circ}$ (Figures~\ref{fig1}c and \ref{fig1}d) and $-60^{\circ}$ (Figures~\ref{fig1}e and \ref{fig1}f), representing S-shaped, toroid-shaped and Z-shaped morphologies, respectively. More details of the implementation process of the RBSL flux rope model in COCONUT can be found in \citet{Guojh2024b}.

\subsection{Modelling the propagation of CME flux ropes with COCONUT}

Once the flux rope is constructed, its subsequent eruption and propagation processes are simulated by COCONUT. The latter is a numerical model which is built upon the Computational Object-Oriented Libraries for Fluid Dynamics (COOLFluiD) platform \citep{Kimpe2005, Lani2005, Lani2013, Lani2014}, employing a fully implicit Finite Volume method to solve full 3D MHD equations on unstructured grids. The adoption of the implicit scheme and unstructured grids facilitates the rapid convergence of steady-state solutions within a domain that encompasses the entire global corona, also including the two poles. Consistent with our previous works \citep{Linan2023, Guojh2024b}, we adopt a polytropic model to simulate the propagation of the CME flux rope from the solar surface to a distance of 25$R_{\odot}$. This model takes into account a reduced adiabatic index $\gamma$ of 1.05, emulating quasi-isothermal heating with limited energy injection \citep{Mikic1999, PerriLeitner2022}, which has demonstrated efficiency in reproducing the large-scale coronal magnetic fields and the propagation of CMEs with COCONUT \citep{PerriLeitner2022, Kuzma2023, Linan2023}. Despite the simplified thermodynamics which is tackled with this model, the evolution of the magnetic configuration remains relatively realistic.

The performance of the COCONUT model involves three main steps. First, we construct the background solar wind using the time-independent relaxation module for the steady-state solution of the background solar wind \citep{PerriLeitner2022}. In this work, the background large-scale magnetic fields and the solar wind are based on the solar magnetogram ("hmi.Synoptic\_Mr\_720s\_small") on 2019 July 2 provided by the Helioseismic and Magnetic Imager \citep[HMI;][]{Scherrer2012} on board the Solar Dynamics Observatory \citep[SDO;][]{Pesnell2012}, for two reasons. On the one hand, this data corresponds to the minimum of solar activity, thereby minimizing the impacts of the background solar wind on CME propagation while still providing a relatively realistic coronal magnetic field environment compared to the dipole configuration. On the other hand, the simple magnetic configuration in the solar minimum activity enhances the stability of the convergence process. In particular, we do not use the raw HMI magnetogram directly as the inner boundary. Instead, the original map is processed using the spherical harmonic projection with a maximum reconstruction frequency of $l_{\rm max}=20$. This preprocessing reduces the strength of the magnetic field, corrects some null values, and smooths small-scale structures. For more details on the preprocessing of input magnetograms, please refer to \citet{Kuzma2023} and \citet{Perri2023}. Additionally, the solar wind configuration which has been reconstructed from this data has undergone extensive examination in our previous COCONUT papers \citep{Kuzma2023, Linan2023}. Next, we superpose the constructed flux ropes into the relaxed solar wind to propel CMEs. Finally, we switch the steady-state relaxation run to a time-dependent simulation in order to study the eruption and propagation process of CMEs. It should also be noted that, our flux ropes are inserted into the solar quiet regions (see Figure~\ref{fig1}), which means that our simulation results mainly concentrate on the CMEs erupting from quiescent filaments. \citet{Linan2023} and \citet{Guojh2024b} provided more details about the CME module in COCONUT.

Our simulation encompasses the entire global corona including both poles, where $r$ spans from the solar surface to a distance of 25$R_{\odot}$ in radius, $\phi$ ranges from $-180^{\circ}$ to $180^{\circ}$ in longitude and $\theta$ ranges from $-90^{\circ}$ to $90^{\circ}$ in latitude. The computational domain is discretized with a 6-level subdivision of a geodesic polyhedron, resulting in approximately 1.5 million cells (see \citeauthor{Brchnelova2022}~\citeyear{Brchnelova2022} for more details). The boundary conditions are prescribed using the approach described in \citet{Linan2023} and \citet{Guojh2024b}. Specifically, at the inner boundary, the radial magnetic field is constrained by the observed magnetogram, while other magnetic field components are determined by zero-gradient extrapolation. The pressure and density are fixed at values of $p_{\odot} = 4.15 \times 10^{-2}\;$dyn cm$^{-2}$ and $\rho_{\odot} = 1.67 \times 10^{-16}\;$g cm$^{-3}$, respectively. The atmosphere of our model is set from the solar corona, and the effects of the photosphere, chrmosphere and transition region on the solar wind and CME propagation are omitted. To eliminate spurious electric fields near the solar surface, the velocity direction is prescribed along magnetic field lines \citep{Brchnelova2022b}. The outer boundary relies upon a zero-gradient extrapolation under the assumption of a super-fast outflow.

\subsection{Dataset for parametric survey}

The input parameters for the implemented flux ropes are summarized in Table~\ref{tab1}. Cases labeled with ``S'', ``Z'' and ``T'' symbols represent flux ropes with S-shaped, Z-shaped, and toroid-shaped projection shapes, respectively. The variation in the projection shape is achieved by changing the value of $\theta$. Cases labeled with ``1'', ``2'' and ``3'' subscripts correspond to the benchmark, cases with a large cross-section radius, and cases with larger toroidal fluxes, respectively. It is worth noting that increasing the cross-section radius of the flux rope inadvertently decrease its magnetic-field strength if keeping the toroidal flux constant. In order to compensate for the magnetic field strength change resulting from the cross-section radius change, we employ a strategy of simultaneously increasing both the cross-section radius and toroidal flux in cases S2, T2 and Z2.

\begin{figure}
  \includegraphics[width=8cm,clip]{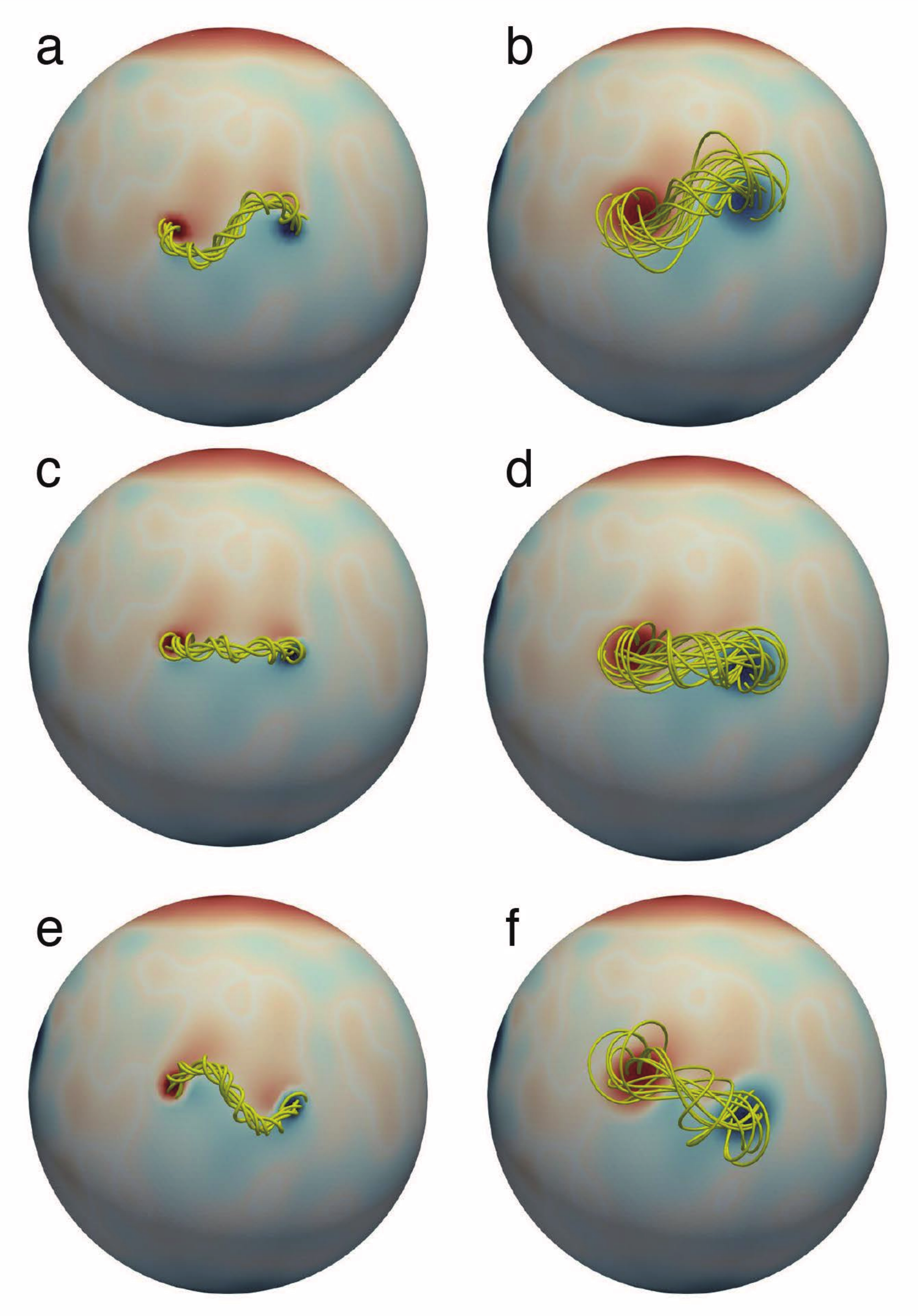}
  \centering
  \caption{Magnetic structures of flux ropes in typical cases. Panels a--f show the cases of S1, S2, T1, T2, Z1 and Z2, respectively. The yellow lines represent flux ropes, which are traced from the flux rope footpoints. The left and right panels depict flux ropes with the small and large cross-section radius, respectively. \label{fig1}}
\end{figure}

Figure~\ref{fig1} displays the reconstructed flux ropes with different projection shapes. The panels from top to bottom show S-shaped, toroid-shaped and Z-shaped flux ropes, respectively. The left and right panels depict slender (small cross-section radius) and fat (large cross-section radius) flux ropes, respectively. It is evident that slender flux ropes are more twisted compared to fat ones, consistent with previous results showing that the twist number of a flux rope increases with its aspect ratio \citep{Wang2016, Guojh2021}. In particular, typically, filaments with positive helicity exhibit S-shaped morphology. Despite this, certain cases may deviate from this empirical rule \citep{Zhou2020}, suggesting that the ``Z'' cases in our dataset are also plausible.

The effects of the preexisting flux ropes on their ensuing CMEs can be elucidated through comparative studies. For example, comparing cases labeled as ``S'', ``Z'' and ``T'' allows us to understand the influence of the projection shape of flux ropes. The comparison between cases denoted by ``1'' and ``2'' subscripts enables us to assess the effect of the cross-section radius, while comparing cases labeled with ``2'' and ``3'' subscripts illustrates how the toroidal flux of flux ropes influences their eruption and propagation processes. This paper mainly concentrates on the propagation and large-scale magnetic structures of CMEs in the outer corona rather than their initiation processes. The inserted flux ropes are in non-equilibrium and are torus-unstable. As a result, CME flux ropes will directly erupt and be accelerated to a high speed upon insertion.

\begin{table*}[h]
\centering
\caption{Parameters and results in different cases. The arguments of $\theta$, $F$, $a$ and $V_{\rm CME}$ denote the writhe angle, toroidal flux, cross-section radius of the flux rope, and the propagation speed of its consequent CME, respectively. \label{tab1}}
\begin{tabular}{lcccccl}
\hline \hline
{}&
Case&
$\theta$&
$F$&
$a$&
$V_{_{CME}}$\\
\hline
{}&
{}&
($^{\circ}$)&
($\times10^{20}$ Mx)&
(Mm)&
(km s$^{-1}$)
\\ \hline
{}& S1 & 60 & 3 & 35 & 465 &  \\
{S-shaped}& S2 & 60 & 12 & 105 & 467 &  \\
{}& S3 & 60 & 24 & 105 & 697 &  \\
\hline
{} & Z1 & -60 & 3  & 35 & 504 &  \\
{Z-shaped}  & Z2 & -60 & 12 & 105 & 456 &  \\
{}  & Z3 & -60 & 24 & 105 & 708 &  \\
\hline
{} & T1 & 0 & 3 & 35 & 484 &  \\
{Toroid-shaped} & T2 & 0 & 12 & 105 & 445 &  \\
{} & T3 & 0 & 24 & 105 & 702 &   \\
{}  & T4 & 0 & 3 & 105 & -- &  \\
\hline \hline
\end{tabular}\label{table1}
\end{table*}

\section{Numerical results}\label{sec:res}

\subsection{Impacts on magnetic structure and propagation speed of CMEs}

\begin{figure}
  \includegraphics[width=8cm,clip]{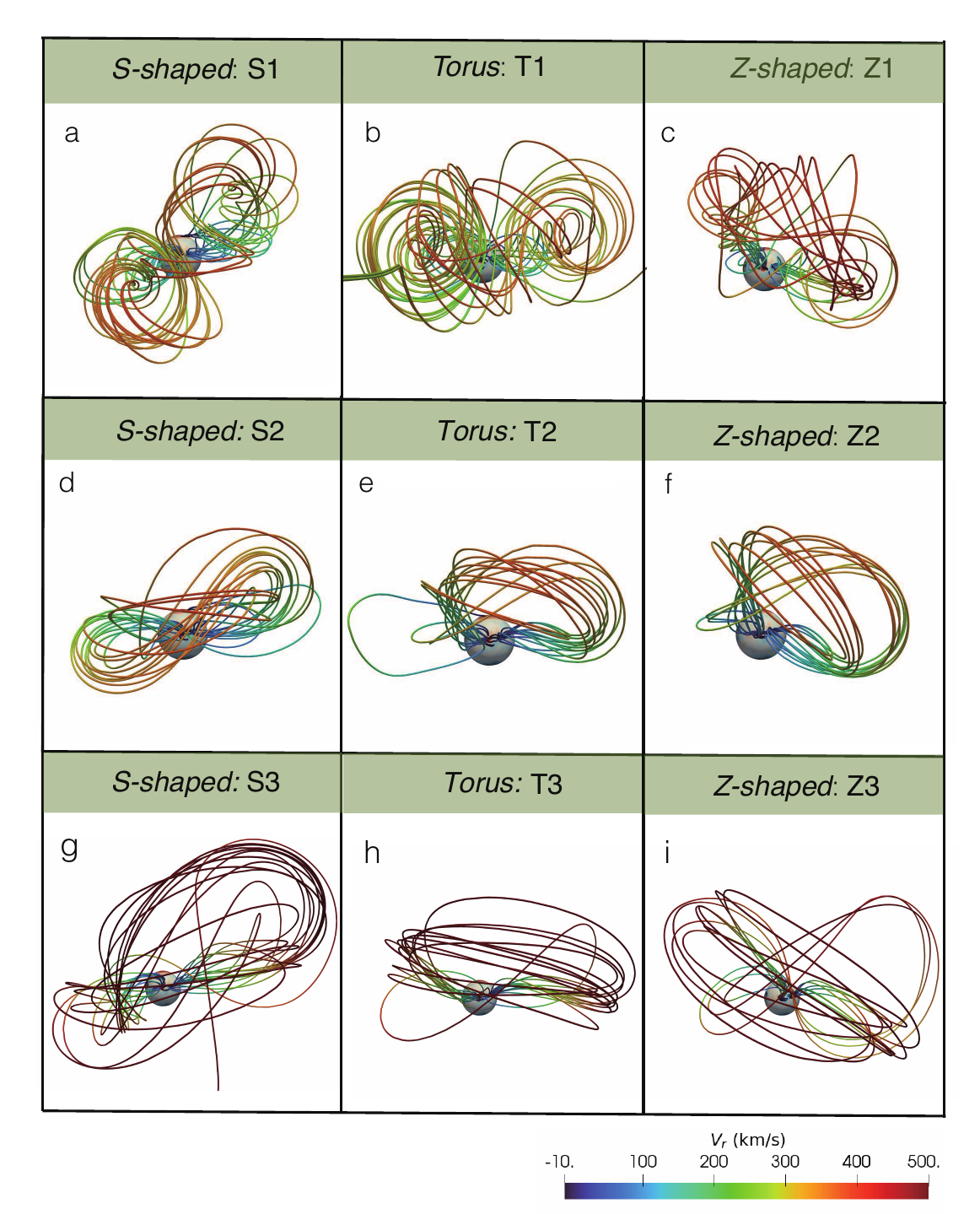}
  \centering
  \caption{Top views of CME magnetic structures at 3.2~hr. The field lines are traced from the locations of the flux-rope footpoints at the beginning. The color represents the radial velocity, which is ranging from -10 to 500~km/s.}\label{fig2}
\end{figure}

\begin{figure}
  \includegraphics[width=8cm,clip]{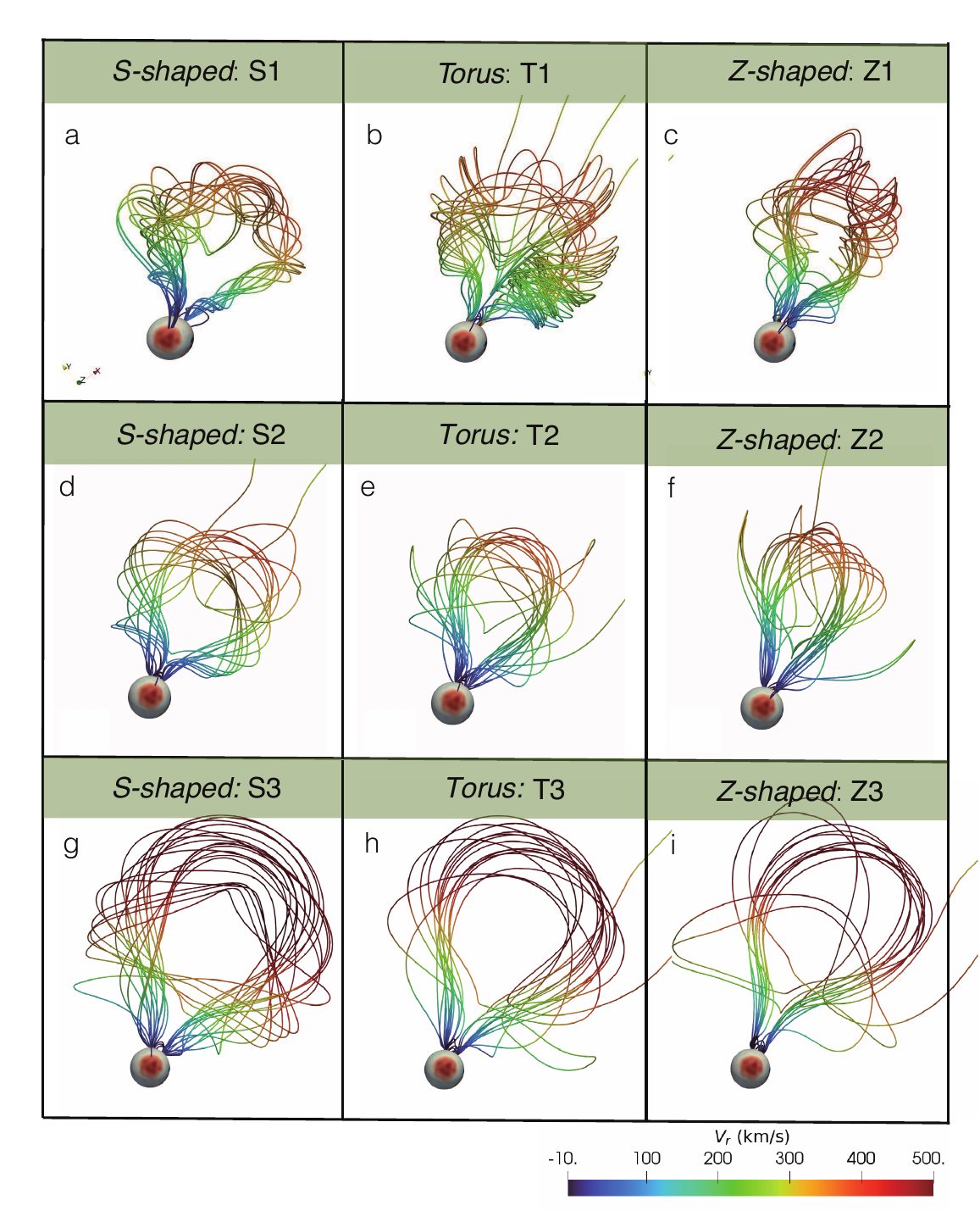}
  \centering
  \caption{The same as Figure~\ref{fig2} but for the side view. \label{fig3}}
\end{figure}

First, we investigate how the morphology of the preexisting flux ropes influences the magnetic structures of their resulting CMEs. Figures~\ref{fig2} and \ref{fig3} present typical field lines of CME flux ropes 3.2~hr after the eruption, traced from the footpoints of the flux rope. It is seen that CMEs retain certain features of their progenitors in many senses, even though they have propagated to a distance far beyond 10$R_{\odot}$. For example, CME flux ropes inherit their initial morphology to a great extent, maintaining the sigmoid (Figure~\ref{fig2}a), toroid (Figure~\ref{fig2}b) and reversed sigmoid (Figure~\ref{fig2}c). On top of that, the size or thickness of the preexisting flux ropes is crucial in determining the twist properties of CME flux ropes. The CMEs originating from slender flux ropes are more twisted than those erupting from fat ones, which is consistent with the findings for CME progenitors \citep{Guojh2021} and CME consequences in interplanetary space \citep{Wang2016}. Moreover, it seems that the bulk-up of initial flux ropes can mitigate the impacts of the projection shape on CME structures. To sum up, our results strongly indicate that both the projection shape and thickness of source flux ropes are worthy of consideration when predicting the magnetic structures of CMEs in the outer corona.

\begin{figure}
  \includegraphics[width=8cm,clip]{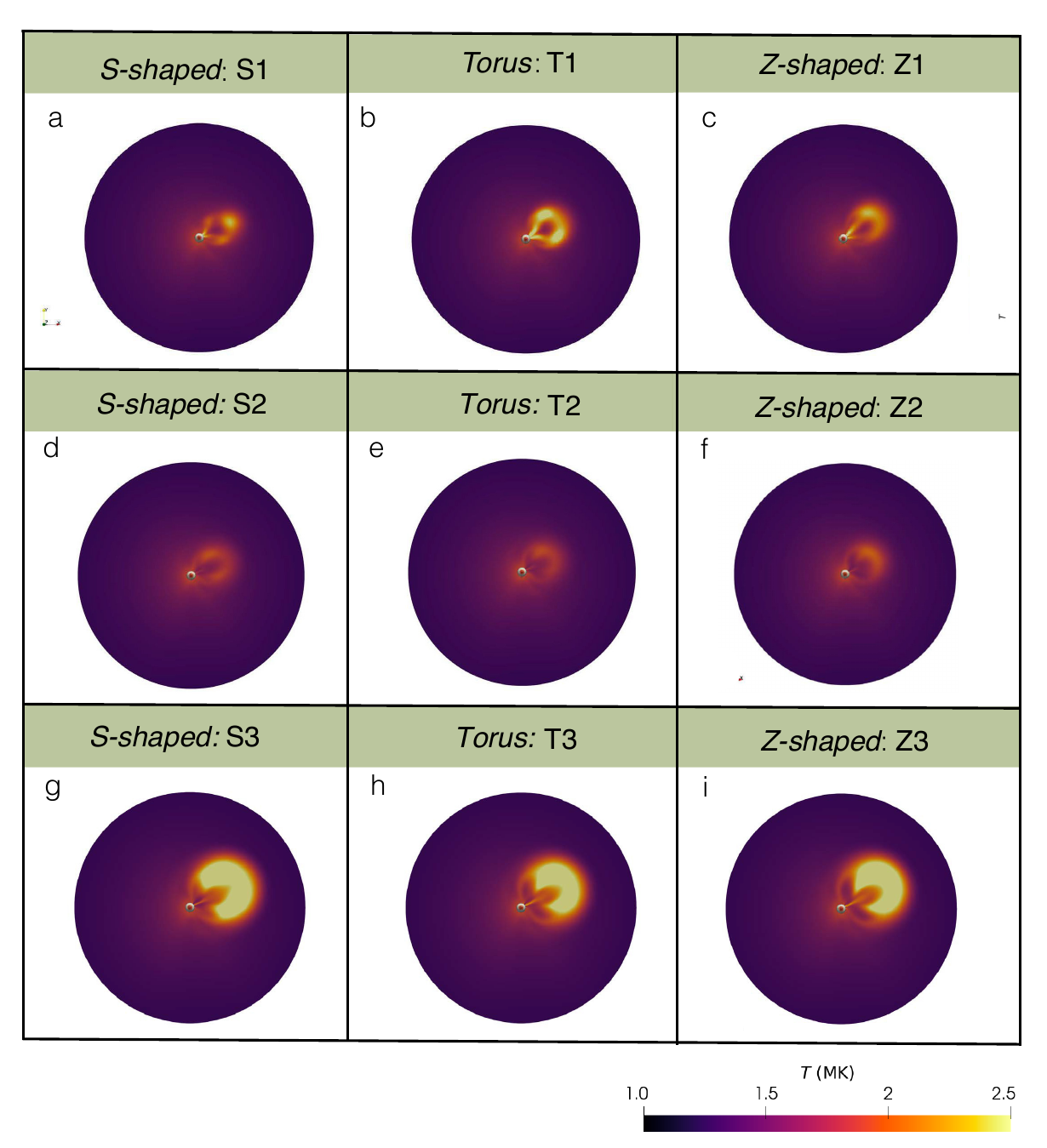}
  \centering
  \caption{Temperature distributions on the equatorial plane at 3.2~hr. \label{fig4}}
\end{figure}

\begin{figure}
  \includegraphics[width=8cm,clip]{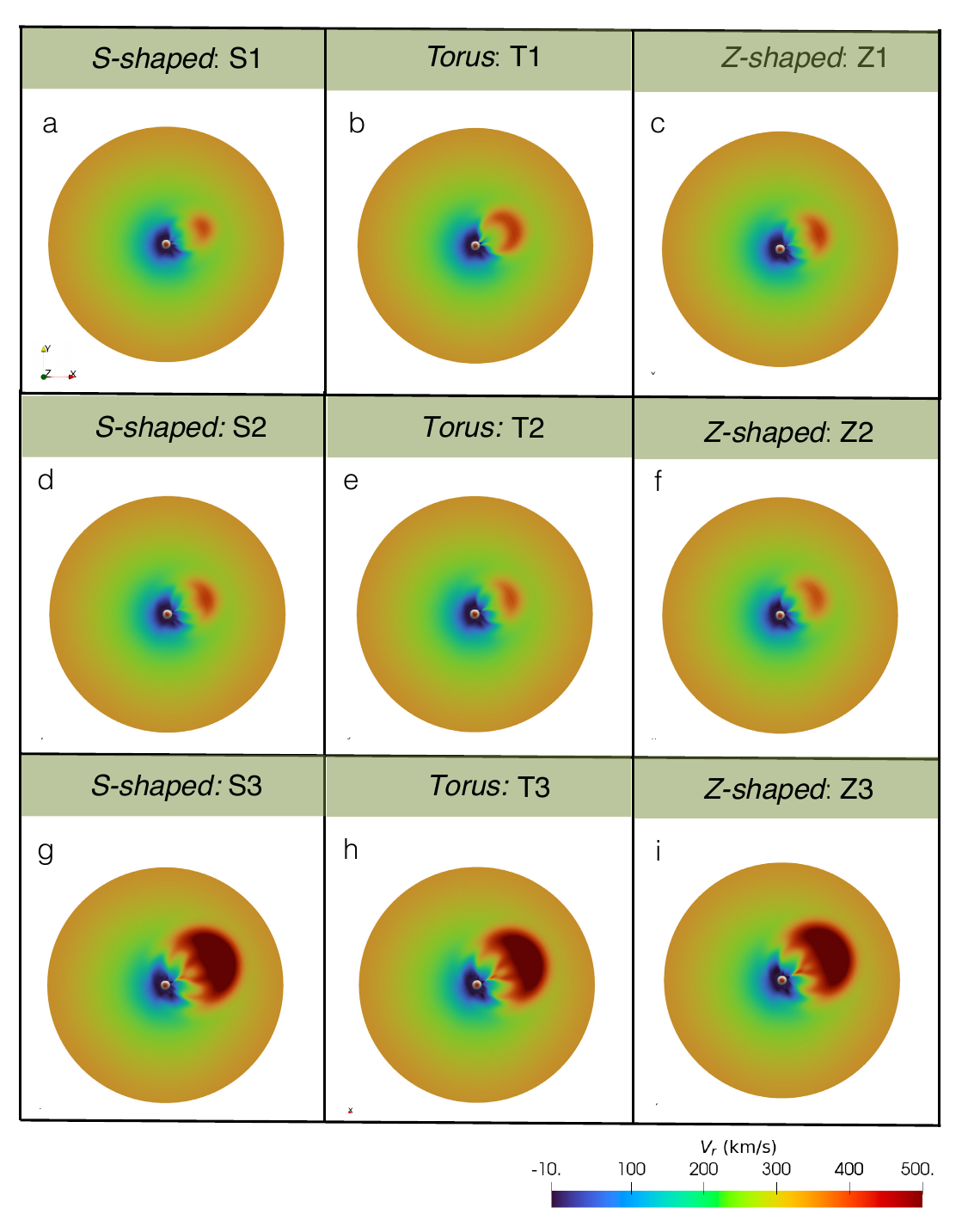}
  \centering
  \caption{Radial velocity distributions on the equatorial plane at 3.2~hr. \label{fig5}}
\end{figure}

In the following, we explore impacts on the thermodynamics of CMEs. Figures~\ref{fig4} and \ref{fig5} illustrate the temperature and radial velocity distribution on the equatorial plane, respectively, from which one can see that CMEs are nearly outlined by the heated areas, driving forward high-speed flows. Furthermore, it is seen that an increase in the toroidal flux of source flux ropes results in faster and hotter CMEs. However, the variation due to the thickness/cross-section radius of source flux ropes is relatively smaller. In particular, it is found that the CME temperature is smaller when simultaneously increasing the toroidal flux and expanding the cross-section radius of flux ropes (by comparing cases T1/T2, S1/S2 and Z1/Z2). This is due to the decrease in electric current density when increasing the volume of the flux rope bulk. Of special note is that the hot and upward flows below flux ropes predicted in the stand flare model \citep{Carmichael1964, Sturrock1966, Hirayama1974, Kopp1976} can be identified, especially in cases S3, T3 and Z3, which is generally regarded as the result of magnetic reconnection. Therefore, magnetic reconnection could be more pronounced in CMEs erupting from flux ropes with high toroidal fluxes over the same propagation duration.

\begin{figure}
  \includegraphics[width=8cm,clip]{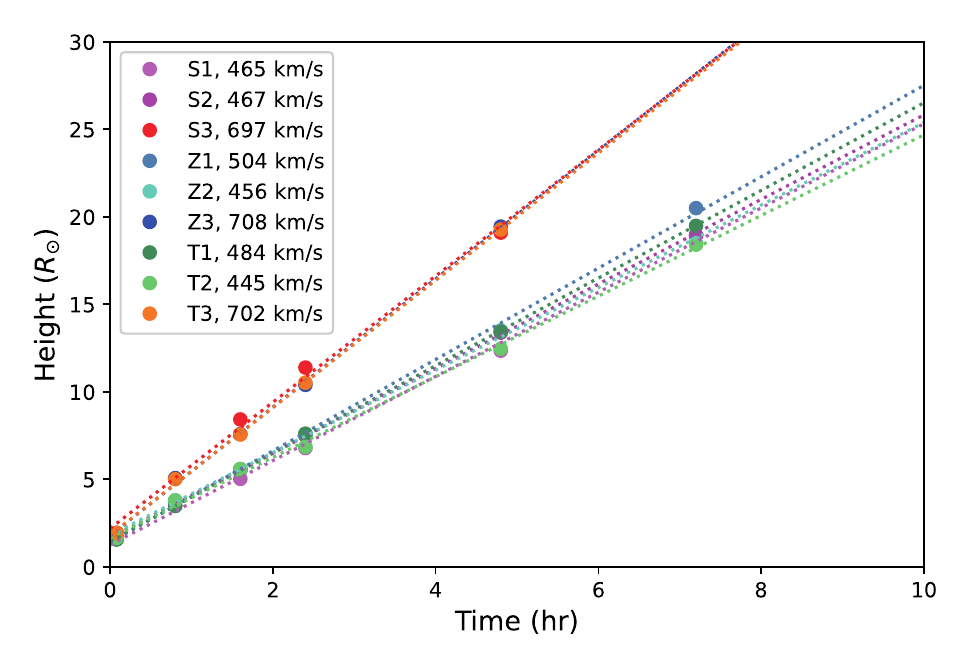}
  \caption{Time-distance diagram of CMEs. The dotted lines are fitted with a linear trend.  \label{fig6}}
\end{figure}

In order to quantify the propagation speed of CMEs, we present time-distance diagrams in Figure~\ref{fig6}, by tracking the CME leading edges measured from the radial direction of the eruption. As a result, the propagation speed of CMEs can be estimated through linear fitting. It is found that CMEs erupting from flux ropes with a large toroidal flux (Cases S3, Z3, and T3) propagate faster compared to the other cases. However, the impact of its projection shape is not apparent. This suggests that an effective approach to propel fast CMEs is by increasing the toroidal flux of source flux ropes, which is consistent with the results of \citet{Linan2023}.

\subsection{Impacts on in-situ detection of magnetic clouds}

\begin{figure*}
  \includegraphics[width=15cm,clip]{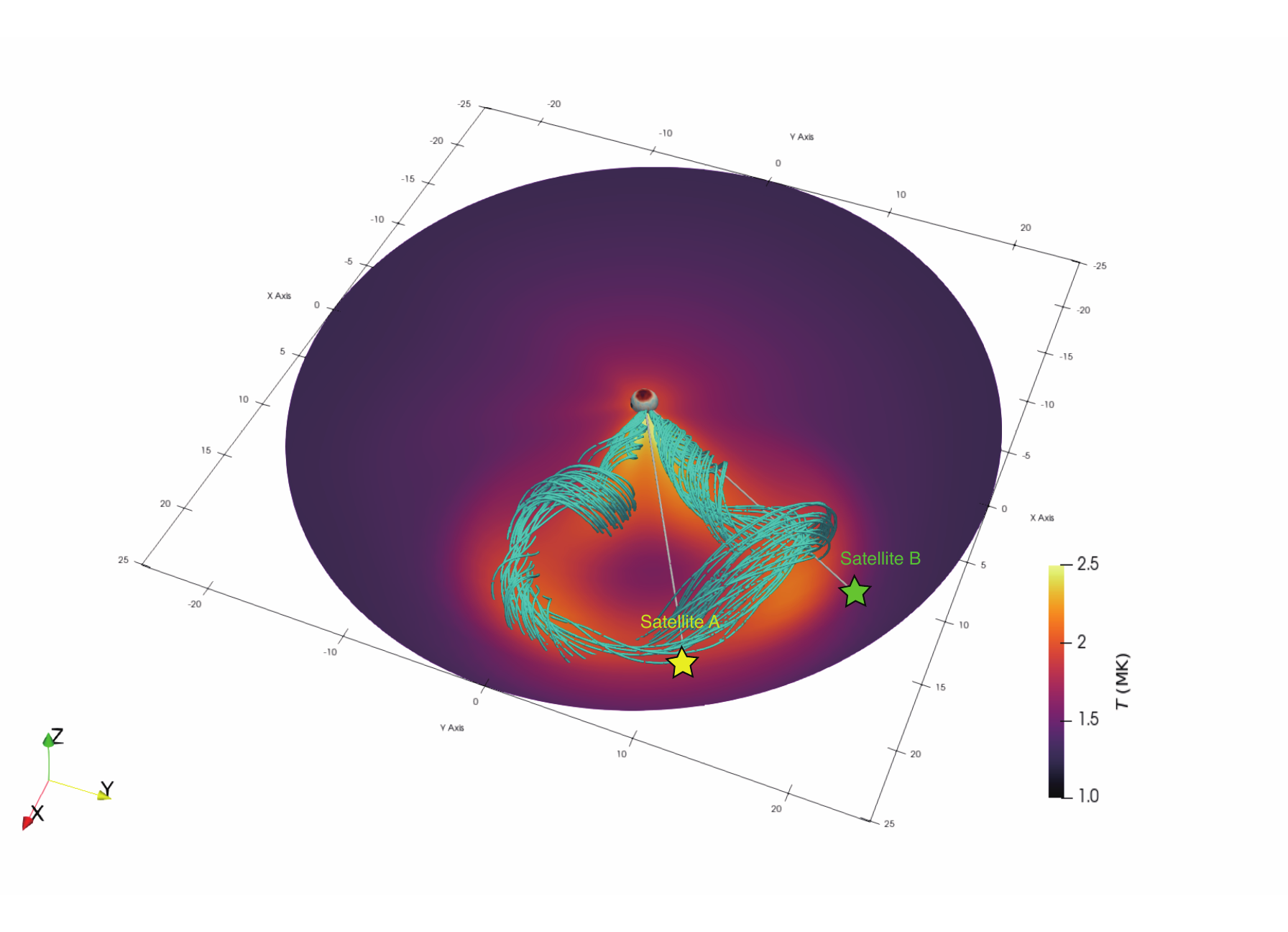}
  \centering
  \caption{Positions of two virtual satellites, in which satellites A and B passes through the nose and flank of modeled CMEs, respectively. \label{satellite_sketch}}
\end{figure*}

\begin{figure*}
  \includegraphics[width=18cm,clip]{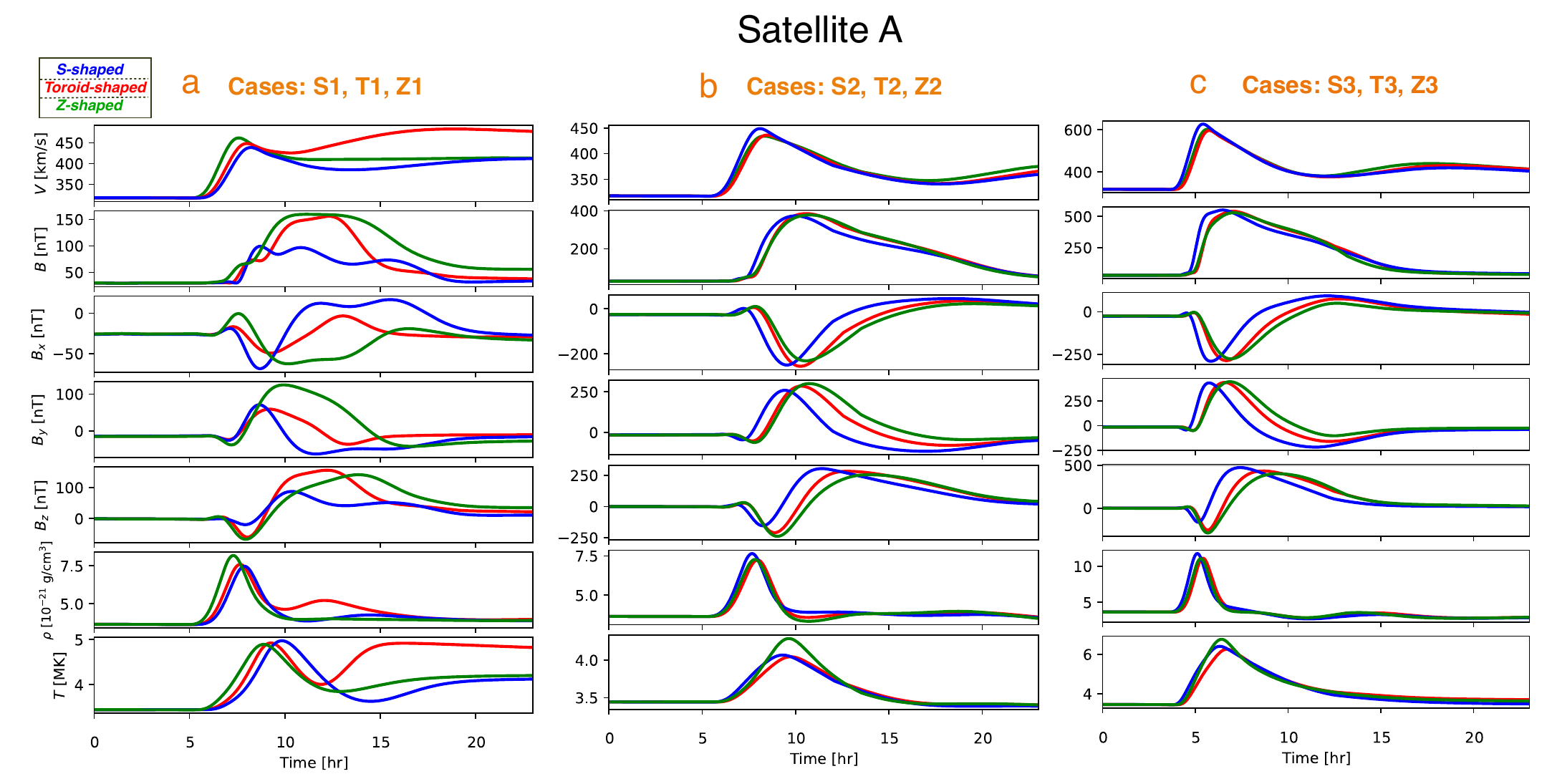}
  \centering
  \caption{In-situ plasma profiles measured at satellite A. Panels~(a), (b) and (c) present the cases  with ``1'', ``2'' and ``3'' subscripts, respectively. The blue, red and green lines represent the S-shaped, toroid-shaped and Z-shaped cases, respectively. \label{fig7}}
\end{figure*}

\begin{figure*}
  \includegraphics[width=18.5cm,clip]{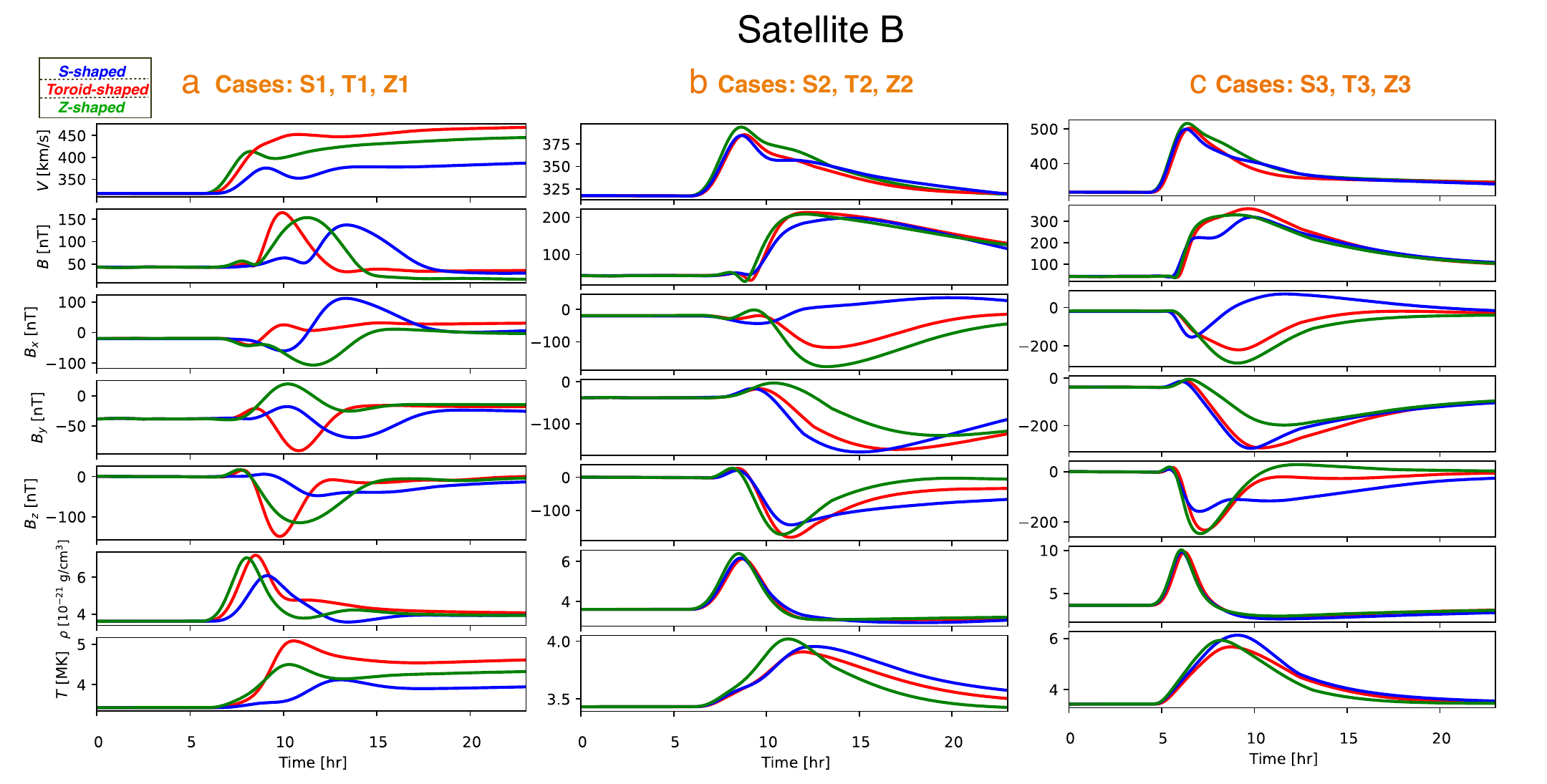}
  \centering
  \caption{In-situ plasma profiles measured at satellite B. \label{saB}}
\end{figure*}

In this subsection, we delve into an exploration of how source flux ropes influence the in-situ plasma profiles of their resulting CMEs. In-situ measurements provide greater precision in diagnosing the local characteristics of CME plasma when compared to remote-sensing observations. To this end, two virtual spacecraft are placed at a heliocentric distance of 21.5 $R_{\odot}$ and are designed to pass roughly through the nose and flank of modeled CMEs, in which satellite A passes through the CME nose, while satellite B mainly detects the features of the CME flank, as shown in Figure~\ref{satellite_sketch}. It is noteworthy that the distance of 21.5$R_{\odot}$ serves as the inner boundary of our heliosphere MHD model, EUHFORIA \citep{Pomoell2019, Poedts2020}. Figures~\ref{fig7} and \ref{saB} illustrate the detected plasma profiles in terms of speed, vector magnetic fields, density, and temperature at satellites A and B, respectively. 

Figure~\ref{fig7}a presents the in-situ profiles of cases S1, T1 and Z1 measured at satellite A (CME nose), in which the detected CMEs result from the eruption of slender flux ropes. It is well seen that the projection shape of source flux ropes leads to a significant change in the plasma profiles of CMEs, especially for magnetic fields. For instance, the CMEs resulting from the eruptions of Z-shaped and toroid-shaped flux ropes exhibit a stronger magnetic-field amplitude compared to the S-shaped case. This may be due to the enhanced magnetic reconnection between S-shaped CME flux ropes and overlying magnetic fields, as suggested by the magnetic connectivity in Figure~\ref{fig1}. Additionally, the Z-shaped case shows a larger negative $B_{z}$ component than the S-shaped case, which is attributed to the fact that Z-shaped flux ropes with positive magnetic helicity can carry stronger negative $B_{z}$ magnetic fields in principle, as shown in Figure~\ref{fig1}.  

Regarding the velocity profiles, despite the similarity in the arrival time of these CMEs, their profiles have obvious difference. A late-phase in case T1 (red line), which is characterized by the heating and acceleration, is observed. Similar phenomena were also found in observations from radio occultation measurements with the Akatsuki spacecraft \citep{Ando2015}. This could be explained by the magnetic reconnection taking place below the flux rope, consistent with the results in Figure~\ref{fig3}, where the highly curved field lines and underlying flare loops serve as evidence of magnetic reconnection \citep{Guojh2024b}. It is worth noting that the late phase is not significantly visible in the toroid-shaped flux ropes with large cross-section radii (cases T2 and T3), which may be due to the different dominant reconnection geometries in these cases. Although the flux ropes in cases T1, T2 and T3 are similar in geometry, they differ significantly in twist number, with the slender flux rope in case T1 being more twisted than fat ones in cases T2 and T3. As a result, separatrices are more likely to exist between CME flux ropes and ambient arcades in case T1, enhancing the interchange magnetic reconnection that flux ropes may participate in. Additionally, the flux ropes with the large cross-section radius occupy a portion of volume originally belonging to the overlying background magnetic fields, thereby reducing the likelihood of magnetic reconnection in the overlying arcade field lines. Instead, the reconnection between two flux-rope legs could be more dominant. This may result in the late phase being less visible in cases T2 and T3 compared to case T1.

Nevertheless, the influences from the morphology are highly sensitive to the cross-section radius of the flux rope. As depicted in Figure~\ref{fig7}b, the in-situ profiles of CMEs which are initiated from fat flux ropes exhibit greater similarity, regardless of the various projection shapes of the preexisting flux ropes. Hence, our results suggest that the axis path of the preexisting flux ropes becomes less critical in modeling the propagation of their resulting CMEs when the cross-section radius is large enough. In this scenario, the flux-rope model with a simply straight toroidal shape remains sufficiently accurate in simulating the propagation of CMEs, even though observed flux-rope proxies, such as filaments, follow intricate paths. Figure~\ref{fig7}c displays cases with high toroidal fluxes (S3, T3, Z3). In comparison to the results in Figure~\ref{fig7}b, flux ropes with high toroidal fluxes can propel CMEs at higher speeds. 

The above detection primarily reflects the plasma characteristics of the CME nose. To understand the results of the CME flank, we present the in-situ plasma profiles detected by satellite B in Figure~\ref{saB}. Similar to the results measured from the CME nose, the in-situ profiles of CMEs initiated from fat flux ropes still remain similar to a large extent, even when detected at their flanks. Moreover, the differences in profiles resulting from the morphology of the source flux ropes are more pronounced at the CME flanks. This may be due to that the non-radial motions (e.g., the rotation direction) differ significantly depending on the projection morphology of the source flux ropes \citep{Zhou2020}, e.g., S-shaped flux ropes generally rotate clockwise.

\section{Discussions} \label{sec:dis}

\subsection{Is the morphology of source magnetic flux ropes crucial for their consequent coronal mass ejections?}

Although magnetic flux ropes detected in the interplanetary space are often modeled as simple cylindrical twisted flux tubes anchored to the solar surface \citep{Isavnin2016}, it is widely accepted that their progenitors in solar source regions, which are conventionally proxied by sigmoids, hot channels, coronal cavities and filaments, often exhibit complicated and irregular morphology. For instance, sigmoids and hot channels typically display S-shaped or Z-shaped structures \citep{Cheng2017}. Furthermore, prominences often show significant variations in morphology, appearing as clouds, stems, horns, flames and plumes. Their length can vary considerably depending on their magnetic surroundings, ranging from less than 100~Mm in active regions to up to 600~Mm in quiet regions \citep{Tanberg-Hanssen1974, Filippov2000, Mackay2010, Chen2020, Gunar2023}. 
Additionally, the width of quiescent filaments is notably larger compared to active-region types, suggesting potential differences in the cross-section radius of their supporting magnetic flux ropes \citep{Guo2022}. 

Despite the intricate shapes exhibited by these CME progenitors in observations, the majority of numerical models extensively simplify them with regular and simple shapes, such as the spheromark \citep{Kataoka2006, Shiota2010, Verbeke2019}, toroid \citep{Titov2014, Linan2023} and teardrop \citep{Gibson1998, Jin2017}. Does the deviation of modeled flux ropes from observations in the solar source region influence the magnetic structures of their resulting CMEs? Under what conditions can we simplify the source flux rope with the simple and regular morphology (such as a toroid)? The ensuing discussions will answer these queries.

Our simulation results unveil the impact of source flux rope morphology on the resulting CMEs. As depicted in the top panels of Figures~\ref{fig3} and \ref{fig7}a, the CMEs erupting from flux ropes with different projection shapes exhibit distinct magnetic structures and in-situ plasma profiles. Specifically, the CME originating from an S-shaped flux rope displays a weaker negative $B_{z}$ component compared to the Z- and toroid-shaped cases. If CME flux ropes maintain the same orientation in the subsequent propagation process, the stronger geomagnetic effects could be induced in latter cases. However, as the cross-section radius of the flux rope increases, the influences from the flux-rope morphology become less pronounced, as evidenced in Figures~\ref{fig7}b and \ref{fig7}c. This suggests that simplifying a flux rope path with a simple shape to model the propagation of CMEs is reliable when its cross-section radius is sufficiently large. \citet{Guo2022} revealed a spatial relationship between filament materials and their supporting flux ropes based on pseudo-3D simulations for the filament formation: filament materials occupy almost a quarter of the flux rope. Based on this and the parameter survey in this paper, we suggest that the morphology of flux ropes become less important when the ratio between the width and length of their hosting filaments exceeds 0.125. Conversely, it becomes essential to measure the realistic paths of flux ropes from their observed proxies. Our findings strongly underscores the advantage of the RBSL method, capable of precisely modeling a flux rope with an axis of arbitrary shape.

On the other hand, the above results reveal the importance of considering the cross-section radius of the flux rope in future CME prediction tools. Various approaches can be employed to derive these topology parameters. The first method relies on the 3D reconstruction of coronal magnetic fields, such as the NLFFF extrapolation \citep{Guo2016, Guo2019a, Guojh2021} and data-driven models \citep{Cheung2012, Jiang2016, Pomoell2019, Guo2024}, which can directly construct magnetic flux ropes and thus outline key topology parameters. The second involves utilizing observational features to trace flux ropes, such as the filament width, drainage sites, flare ribbons, dimming and sunspot scar \citep{Harra2001, Wangws2017, Aulanier2019, Xing2020, Wang2023, Xing2023}. Therefore, to enhance the accuracy in CME prediction, it is important and urgent to develop novel approaches to diagnose the key topology parameters of source magnetic flux ropes, such as the axis path and cross-section radius. However, it should be noted that, the flux ropes in our dataset are positive in helicity sign and positioned at the solar equator. This implies that the results in real observations may be more complicated than the conclusions drawn in this paper.

\subsection{Linking the CME propagation speed and successfulness with its progenitor}

It is widely accepted that the speeds of CMEs and their associated driven shocks are closely linked to particle acceleration and the induced geomagnetic effects \citep{Chen2011, Webb2012, Tsurutani2023}. Therefore, identifying flux-rope parameters that are sensitive to CME speed is crucial for forecasting the impacts of CMEs on the heliospheric environment.

As shown in Figure~\ref{fig6}, the propagation speed of CMEs is not sensitive to the morphology of their source flux ropes, but significantly increases with the toroidal flux. Similar conclusions concerning the effects of toroidal flux have also been drawn in previous studies. For example, \citet{Qiu2007} found that CME speed increases with the reconnection fluxes identified from flare ribbons. Besides, \citet{Chen2006} found that the CME speed is roughly proportional to the average magnetic strength of the source photosphere. \citet{Su2011} found that an increase in toroidal flux can lead to the catastrophic loss of equilibrium, thereby resulting in an eruption. \citet{Zhang2020} found that flux feeding can cause the flux rope to rise. Additionally, global MHD simulations based on the Gibson-Low flux rope model \citep{Jin2017} and Titov-D{\'e}moulin-modified model \citep[TDm;][]{Linan2023} have also revealed that CMEs erupting from flux ropes with higher toroidal flux tend to propagate more rapidly. On the other hand, the source magnetic flux ropes in cases T1 and T4 have the same toroidal flux, but case T4 has a larger different cross-section radius. The fact that case T4 turns out to be a confined eruption implies that the large size of the source flux rope may hinder the CME from erupting with a high speed (e.g., via a bigger drag force), and may even result in failed eruptions.

\begin{figure*}
  \includegraphics[width=15cm,clip]{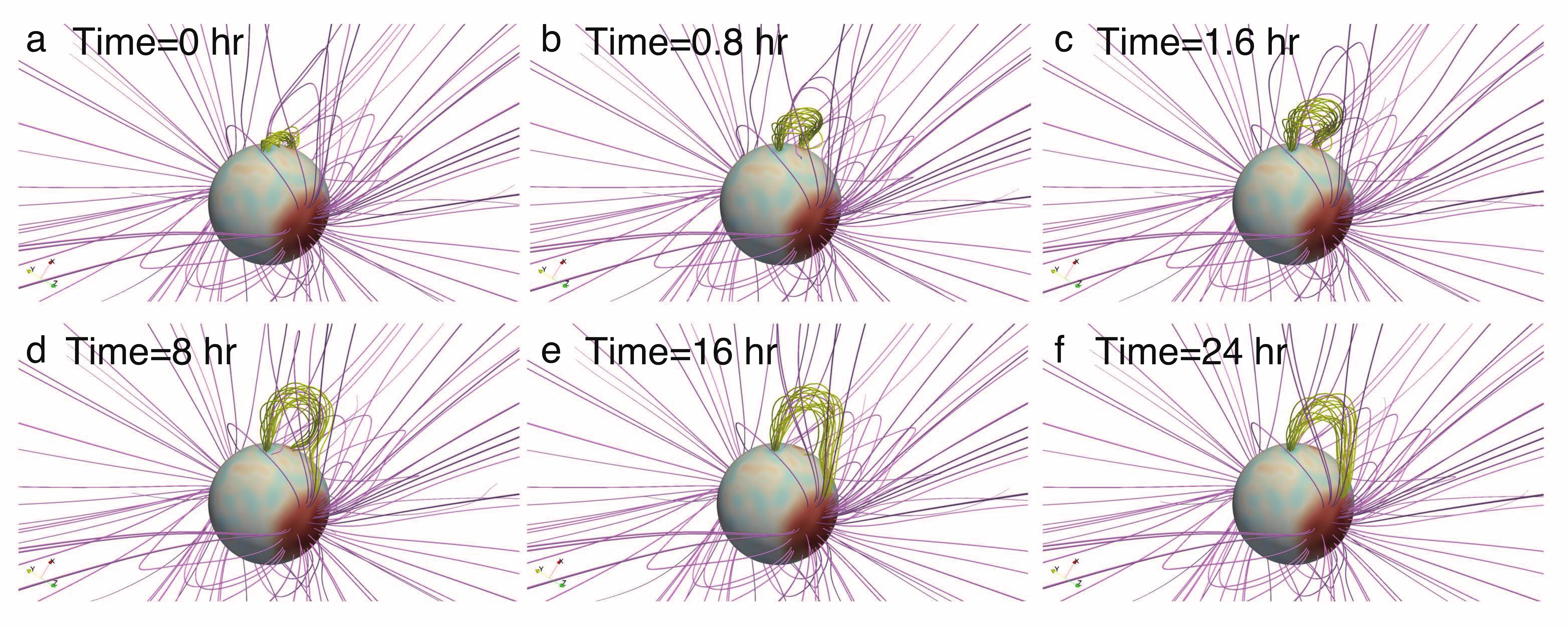}
  \centering
  \caption{The evolution of magnetic fields of a confined eruption (case T4). The yellow and pink lines represent the flux rope and the background solar-wind magnetic fields, respectively. \label{fig8}}
\end{figure*}

Given this insight, we can infer that the eruption may be constrained in the solar atmosphere when the toroidal flux of source flux ropes is sufficiently small. Previous works have demonstrated that strong overlying magnetic fields can constrain solar eruptions, namely, magnetic cage \citep{Amari2018}. However, the properties of the core fields should also be effective in influencing the kinetics of the resulting CMEs. To investigate this, we conduct a simulation, called case T4, in which the toroidal flux is approximately one-eighth of that in case T2, while keeping other parameters constant. The 3D evolution of the magnetic fields in case T4 is illustrated in Figure~\ref{fig8}. Initially, the flux rope undergoes rapid ascent, accompanied by a slight clockwise rotation due to the release of twist. Following this, the rising of the flux rope starts to decelerate, with some twisted field lines turning into expanding loops. Meanwhile, it interacts with neighboring streamers which extend from the polar regions through interchange reconnection. As a result, after one day of evolution, the twisted flux rope disintegrates into the overlying loops, with one of their footpoints migrating toward the polar regions. This case demonstrates that the toroidal flux not only plays a crucial role in determining CME speed but also can determine the success or failure of an eruption. Therefore, our parametric survey suggests that deriving the toroidal flux of pre-eruptive flux rope is significant in CME forecasting, such as the ability to escape from the corona and the speed of the eruption.

\section{Summary} \label{sec:sum}

In this paper, we investigated the impacts of magnetic flux ropes in solar source regions on their resulting CMEs, with the aid of 3D global MHD simulations from the solar surface to a distance of 25$R_{\odot}$. Our parameter survey mainly focuses on the CME events initiated from quiescent filament eruptions in observations. Our parametric survey established a bridge between the large-scale CMEs at 20 $R_{\odot}$ and their solar source regions, leading to the following conclusions:

\begin{enumerate}

\item{The projection shape of source flux ropes is considerable in changing the magnetic structures and in-situ plasma profiles of their resulting CMEs, particularly for CMEs erupting from slender flux ropes. This underscores the importance of accurately measuring the morphology of flux ropes from CME progenitors in observations. However, the impact of the projection shape on the speed of CMEs is not significant.}

\item{The ratio of cross-section radius/thickness of source flux rope is a crucial parameter for CME prediction. We find that the effects arising from the projection shape of source flux ropes on their resulting CMEs can be ignored as their cross-section radius is sufficiently large.}

\item{Under the same overlying magnetic fields, the propagation speed of CMEs increases with the toroidal flux of their source flux ropes. Additionally, it is also found that there is a tendency for confined eruptions to occur when the toroidal flux of the source flux rope is too small.}

\end{enumerate}

The outcomes of this paper offer valuable insights for space weather forecasting. First, they aid in identifying high-risk events with the potential to cause geomagnetic effects from their progenitors in solar source regions. Second, these findings can provide constraints for the input parameters of launched CMEs at 21.5 $R_{\odot}$ in heliospheric simulations, such as with EUHFORIA. Moreover, the bridge between CMEs and their source flux ropes, as revealed by our parametric survey, offers valuable insights for reproducing real eruption events in observations.

\begin{acknowledgements}
This work is supported by the National Key R\&D Program of China (2020YFC2201200, 2022YFF0503004), NSFC (12127901, 12333009), Shanghai Institute of Satellite Engineering (SAST2021-046, D050101), projects C14/19/089  (C1 project Internal Funds KU Leuven), G.0B58.23N and G.0025.23N (WEAVE) (FWO-Vlaanderen), SIDC Data Exploitation (ESA Prodex-12), AFOSR FA9550-18-1-0093, and Belspo project B2/191/P1/SWiM. The resources and services used in this work were provided by the VSC (Flemish Supercomputer Centre), funded by the Research Foundation - Flanders (FWO) and the Flemish Government. 
\end{acknowledgements}

\bibliographystyle{aa}
\bibliography{ms}

\end{document}